# Threat Detection for General Social Engineering Attack Using Machine Learning Techniques

Zuoguang Wang,[1,2,*] Yimo Ren,[1,2] Hongsong Zhu,[1,2,*] and Limin Sun[1,2]

[1] School of Cyber Security, University of Chinese Academy of Sciences, Beijing 100049, China.
[2] Beijing Key Laboratory of IoT Information Security Technology, Institute of Information Engineering, Chinese Academy of Sciences (CAS), Beijing 100093, China.

Correspondence：Zuoguang Wang (wangzuoguang16@mails.ucas.ac.cn) Hongsong Zhu (zhuhongsong@iie.ac.cn)

## Abstract

This paper explores the threat detection for general Social Engineering (SE) attack using Machine Learning (ML) techniques, rather than focusing on or limited to a specific SE attack type, e.g. email phishing. Firstly, this paper processes and obtains more SE threat data from the previous Knowledge Graph (KG), and then extracts different threat features and generates new datasets corresponding with three different feature combinations. Finally, 9 types of ML models are created and trained using the three datasets, respectively, and their performance are compared and analyzed with 27 threat detectors and 270 times of experiments. The experimental results and analyses show that: 1) the ML techniques are feasible in detecting general SE attacks and some ML models are quite effective; ML-based SE threat detection is complementary with KG-based approaches; 2) the generated datasets are usable and the SE domain ontology proposed in previous work can dissect SE attacks and deliver the SE threat features, allowing it to be used as a data model for future research. Besides, more conclusions and analyses about the characteristics of different ML detectors and the datasets are discussed.

## 1. Introduction

### 1.1. Background

Social Engineering (SE) describes a type of attack in which the attacker exploit human vulnerability through social interaction (by means such as influence, persuasion, deception, manipulation and inducing) to breach cyber security (such as confidentiality, integrity, availability, controllability and auditability of infrastructure, data, resource, user and operation in cyberspace) [1]. Social engineering attack includes many types and take various forms such as pretexting, vishing, shoulder surfing, phishing, spear phishing, whaling, baiting, impersonating, smishing, trojan attack and watering-hole. As a type of infamous but quite popular attack, social engineering has posed an increasingly serious security threat. According to reports [2] and [3] from ISACA's State of Cybersecurity, social engineering is the top cyberthreat for organizations from 2016 to 2018. Study [4] shows that social engineering attacks were experienced by 85% of organizations in 2018, an increase of 16% over one year. The average annual cost of social engineering attacks for organizations in 2018 has exceeded $1.4 million, an increase of 8% compared to the previous year [4]. In 2019, email remained the most used attack medium for social engineering attacks, with 88% of organizations worldwide suffering from phishing attacks [5]. 84%, 83%, and 81% of organizations have experienced social engineering attacks such as smishing, vishing, and baiting

respectively [5]. In 2020, more than 80% of organizations in the United States are still subject to various forms of social engineering attacks [6]. Therefore, detecting social engineering attack threat is of great significance for preventing cyber-attacks and ensuring cyberspace security.

### 1.2. Related Work and Gap Analysis

In the area of social engineering threat detection, some studies have made efforts and contributions. Hoeschele and Rogers [7] proposed a decision tree to detect social engineering attacks perpetrated over telephones. It analyzes phone conversations and determine if the caller is deceiving the receiver. However, this method is just a theoretical solution and in the proof-of-concept phase, depending on some unimplemented components such as real-time voice-to-text conversion and analysis subsystem. For call center employees who are easy targets for the social engineering attacks, studies [8] and [9] suggested a social engineering attack detection model which makes use of a textual questions and answers decision tree to server as guidelines to aid call center employees' decision-making. This is a manual, interactive and non-automatic detection approach. Besides, above works aim at phone conversation and are merely applicable to some pretexting and vishing scenarios.

Some works employ machine learning (ML) techniques to detect phishing attacks. Sandouka et al. [10] investigated the feasibility of using neural networks to detect social engineering in the call centers conversation scenarios. But the experiment was simple, in which the features used for training and testing depends on identifying keywords like "install", and only 20 conversation text were used. Basnet et al. [11] compared four machine learning algorithms, i.e. Support Vector Machine (SVM), Biased Support Vector Machines, Neural Network and K-means, for the purpose of classification of phishing emails using features such as HTML email, IP-based URL, age of domain name, number of domains, number of sub-domains, presence of JavaScript. They [11] found that SVM achieved consistently the best results on the selected two datasets. Bhakta and Harris [12] present an approach to detect phishing emails by semantic analysis of dialogs, in which a pre-defined topic blacklist and topic string comparison method was used. Sawa et al. [13] attempt to detect social engineering threat in dialog text by combining Natural Language Processing (NLP) techniques and topic blacklist. NLP is used to detect questions and commands in the text and extract their likely topics. Each extracted topic is compared against a topic blacklist to determine if the question or command is malicious [13]. Peng et al. [14] present an approach to analyze email text and detect inappropriate statements which are indicative of phishing attacks using NLP techniques. This work focused on the natural language text yet didn't rely on analysis of metadata associated with emails, so it is only effective for detecting pure text phishing emails. Lansley et al. [15][16][17] developed a method to detect social engineering attacks in phone conversations based on NLP and Artificial Neural Network (ANN). The conversation text is first parsed and checked for grammatical errors, and then the ANN is used to classify possible attacks. Mridha et al. [18] introduced a ML algorithmic classification models to detect phishing URLs based on Random Forest and ANN. Mondal et al. [19] focused on the spear phishing email detection and found that the ensemble ML approach is more effective than traditional ML techniques. Alsufyani and Alzahrani [20] studied text phishing detection with ML approach and found that K-nearest neighbors (KNN), Decision Tree and AdaBoost [21] models are effective in the detection. Compared with previous works, these studies promote the automatic detection. However, these studies also merely aim at a few of specific social engineering attack scenarios, such as phone conversations, phishing email, phishing URLs, which are not enough or applicable to detect general social engineering attacks[1].

---

[1] General social engineering attacks: attacks that satisfy the definition of social engineering attack, but not limited to a specific social engineering attack type.

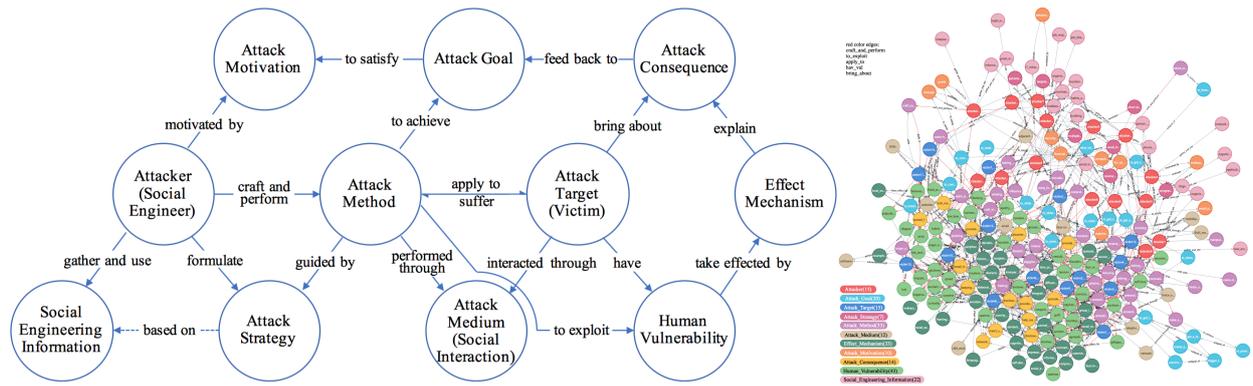

Figure 1: Social engineering domain ontology and knowledge graph proposed in work [22]

Wang et al. [22] proposed a domain ontology for social engineering, which includes 11 types of core entities that significantly constitute or affect the social engineering domain together with 22 kinds of relations among these entities. It provides a formal and explicit description of social engineering attack and an extensible and shareable knowledge schema. This ontology not only models the structure and components of social engineering attacks but also is a further illustration and supplement for the social engineering definition proposed in [1]. The ontology therefore can be used to represent and dissect general social engineering attacks and serve as a framework for social engineering threat analysis and detection. Furthermore, study [22] built a social engineering knowledge graph (covers 14 types of SE attacks) and present four threat analysis methods (analyze individual social engineering attack scenarios or incidents; analyze the most exploited human vulnerabilities; analyze the most used attack mediums; analyze the same-origin attacks) and three potential threat detection methods (detect potential threat, i.e. attackers and attack methods, for victims; detect potential attack methods and targets for attacker; detect potential attack paths from specific attacker to specific target) for general social engineering attacks. The experimental results show that the four threat analysis methods can effectively extract the threat elements, and the F1-scores of the three potential threat detection methods all exceed 0.99. Although these graph-based social engineering threat analysis and detection approaches has obvious advantages in flexibility and visualization, some manual work is concomitant and the detection efficiency relies on the completeness of knowledge graph to some extent.

### 1.3. Motivation and Goal

Therefore, this paper aims to explore the feasibility of different machine learning models for general social engineering threat detection, which on one hand serve as a technical supplement to work [22], i.e. extracting SE threat features and training ML classifiers to learn the inherent laws of datasets to determine whether an attacker poses a SE threat to a target rather than depending on the logical rules, path accessibility or sub-graph retrieval in knowledge graph. On the other hand, to serve as a way 1) to further apply and analyze the SE knowledge graph dataset [22], 2) to examine the effectiveness of the proposed ontology [22] in modeling the inherent features of SE attacks, and 3) to compare the performance difference between knowledge graph-based and ML-based SE threat detection techniques.

### 1.4. Contribution

This paper makes the following contributions.
- This paper extracted different SE threat features and generated new datasets for ML techniques to detect general SE threat, based on yet beyond the labeled knowledge graph dataset in [22]. Experiments and analysis demonstrated the usability of these datasets.

3This work is licensed under a Creative Commons Attribution-NonCommercial-NoDerivatives 4.0 License.
For more information, see https://creativecommons.org/licenses/by-nc-nd/4.0/

- This paper verified the feasibility and effectiveness of ML techniques to detect general SE threat, by training 9 types of machine learning models based on 3 kinds of feature combination datasets respectively and comparing their performance.
- This paper further showed that the proposed SE domain ontology in [22] can effectively dissect SE attacks and model inherent SE threat features, and it therefore can be used as a Data Model to build larger SE datasets used for more security studies on social engineering attacks.

## 2. Method

The method of threat detection for general SE attack consists of four main steps, i.e. data preprocessing, feature extraction and fusion, model training, and threat detection. Python programming language and its libraries are used to implement these tasks.

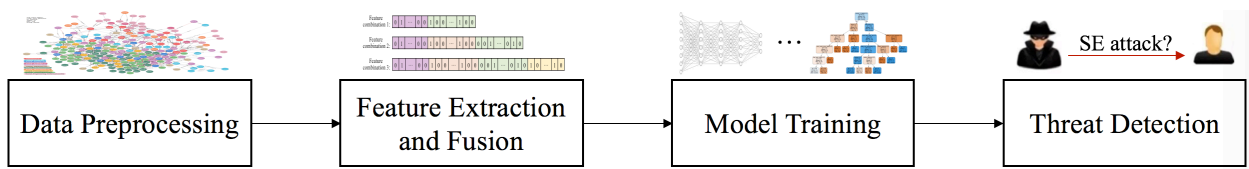

Figure 2: The framework of threat detection for general SE attack using ML techniques

### 2.1. Data Preprocessing

The purpose of this section is to train ML models and then to determine whether an attacker can succeed in performing a SE attack (i.e. posing a SE attack threat) towards a certain target, which requires the preparation of data that characterizes the attackers and victims, as well as label data that indicates the threat state (i.e. posed a SE threat or not).

Table 1: Basic data and information after data preprocessing

| No. | Basic Data and Information | Number |
|---|---|---|
| 1 | attackers | 15 |
| 2 | victims / targets | 15 |
| 3 | attack methods | 33 |
| 4 | human vulnerabilities | 43 |
| 5 | attack mediums | 12 |
| 6 | effect mechanisms | 33 |
| 7 | the use information of attack methods by attackers | 25 |
| 8 | the victims' human vulnerability | 88 |
| 9 | the exploit information of the human vulnerability by attack methods | 97 |
| 10 | the use information of attack mediums by attack methods | 29 |

Study [22] built a SE knowledge graph dataset that covers 14 types of social engineering attacks, containing 1785 triples, 939 relations, and 344 resource nodes. Based on this graph dataset, this paper extracted and preprocessed more information, including the list of *attack methods*, the list of *human vulnerabilities* [23], the list of *attack mediums*, the list of *effect mechanisms* [23], the information about *what attack methods the attackers craft and perform*, the *victims' human vulnerability* [22][23], the *exploit information of the human vulnerability by attack methods*, etc. As Table 1 shows.

Then, we processed the labels of positive samples (posed SE threat) and negative samples (did not pose SE threat). Finally, 225 labels are generated from Table 1, including 138 positive sample labels and 87 negative sample labels, and the former accounts for 61.3% (the baseline precision, recall and F1-score of the positive samples). As Table 2 shows.

Table 2: Label data used in the training dataset

| Data Items | Description |
|---|---|
| labels number | 225 |
| positive samples | 138 |
| negative samples | 87 |
| baseline precision of the positive samples | 61.3% |

**2.2. Feature Extraction and Fusion**

After the preliminary analysis and testing, we focus on four types of features that are closely related to the occurrence of SE attacks.

- Attack methods feature: the attack methods crafted and performed by the attackers reflect the attackers' ability. The stronger the attackers' ability, the more likely the attack will occur.
- Human vulnerabilities feature: the victims' vulnerabilities reflect their potential to be attacked by social engineering. The more typical and numerous the target's vulnerability, the more likely an attack will occur.
- The features of how attack methods exploit the vulnerabilities (vulnerability feature of SE attack method): different kinds of attack methods have differences in the exploitation of human vulnerabilities in terms of categories and quantities, which also indirectly describe the attackers' features.
- Attack mediums: social engineering attacks occur through interactive medium, and the victims' activity and susceptibility differ in different interactive media, which can be used as a feature to describe the victims.

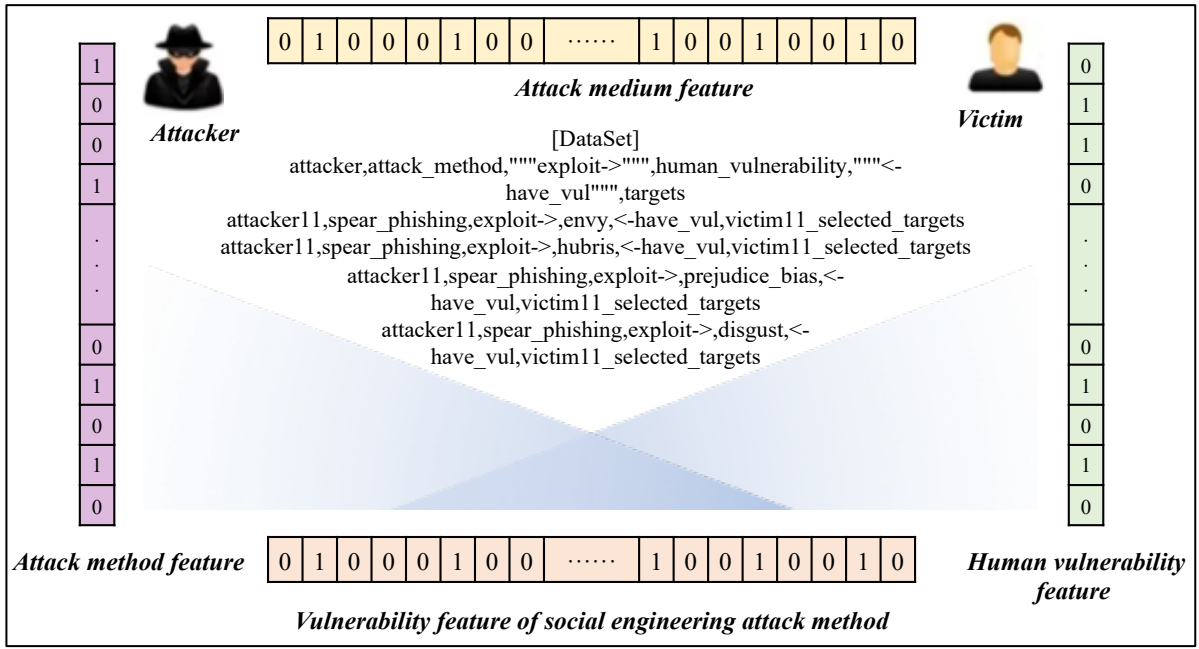

Figure 3: The feature extraction about social engineering threat

This paper employs One-Hot Encoding to vectorize the above features, respectively. There are 33 attack methods in total, and the attackers' attack method feature is represented by a 33-dimensional vector (feature vector 1). If the attacker can craft and perform one attack method, the corresponding status bit of the attack method will be set to 1; otherwise, it will be set to 0. Similarly, the human vulnerability feature of the victims is converted as a 43-dimensional feature vector (feature vector 2), the vulnerability feature of SE attack method is expressed as a 43-dimensional feature vector (feature vector 3)，and the victims' attack medium feature is described as a 12-dimensional feature vector (feature vector 4). As Figure 3 shows.

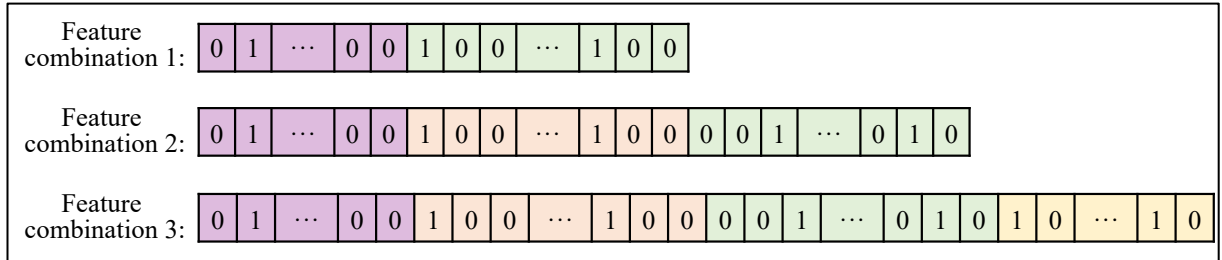

Figure 4: The feature fusion of social engineering threat

In order to analyze the detection performance of the machine learning models under different features conditions, this paper constructs three different feature vector combinations, which make up three training datasets respectively. As Figure 4 shows.

- Feature combination 1 merely considers the attacker's attack method feature and the victim's human vulnerability feature. These two features are relatively independent. The training data is composed of the concatenation of feature vector 1 and feature vector 2, and each piece of training data is a 76-dimensional combined vector (doesn't count the data label).
- Feature combination 2 inserts *the vulnerability feature of SE attack method* into the feature combination 1, which consists of feature vectors 1, 3 and 2, and composes a 119-dimensional combination vector. This combination is essentially the encoding of the attack path described in knowledge graph [22], by which this paper attempts to analyze the learning performance of machine learning methods for logic rules in graph. Feature vector 3 has a certain correlation with feature vector 1. Feature vector 3 is a further description of the attackers' ability, which reflects the logical connection of different features to a certain extent.
- Feature combination 3 considers one more feature than feature combination 2, i.e. the victims' attack mediums feature. It is composed of feature vectors 1, 3, 2 and 4, forming a 131-dimensional combined vector.

Finally, three feature combination datasets (called dataset 1, dataset 2 and dataset 3) are generated for model training and test validation. Since the dimension of the feature vector is not high, we omit the feature dimension reduction for different feature combinations.

**2.3. Model Training**

This section creates 9 ML classification models to train different social engineering threat detection classifiers for the 3 different feature combinations datasets presented in Section 2.2, respectively. These 9 ML classification models are Decision Tree (DT), Random Forest (RF), Support Vector Machine (SVM), Multilayer Perceptron (MLP), Logistic Regression (LR), K-Nearest Neighbor model / Nearest Centroid model, Naive Bayes (NB), Adaptive Boosting (AdaBoost), and Voting ensemble.

Decision Tree (DT) is a supervised learning model, which predicts the classification of target objects by learning a set of decision rules from training data. During the training of the DT classifier, the division impurity represented by the Gini index (equation 1) is used as the standard to determine the growth of the tree, and the attribute with the smallest Gini index under the current conditions is iteratively selected as the child node, so the maximum information gain is obtained on each node. The tree grows until the leaf node's Gini index is 0 or meets certain pre-defined conditions. Usually, the maximum depth of the tree will be set according to the testing process to prevent overfitting (the maximum depth is set to 6 in this paper). Compared with other classification models, the training speed of DT is fast, and the learned decision rules are easy to understand and explain, DT therefore can clearly show which attributes are more important in the classification. In many scenarios, the classification performance of DT is still effective even if the generated set of rules (i.e. DT) is different from the true model of the data.

$$\text{Gini}(D, x) = \sum_{k=1}^{K} p_k(1 - p_k) \tag{1}$$

Random Forest (RF) is an ensemble learning model. First, multiple sub-decision trees are trained by randomly sampling multiple sub-datasets from the training dataset with replacement, and these sub-decision trees form a decision forest. RF output decision result by combining different sub-decision trees and averaging the prediction results of multiple sub-decision trees. Thus, RF mitigates the risk of high variance and overfitting of a single decision tree. RF is also faster to train. Besides, RF can not only learn knowledge from higher dimensional data, but is also helpful to balance the error of the dataset, which is especially suitable for imbalanced datasets. On the other hand, the performance of RF may be limited for data with small size, few features, or low feature dimensionality.

Support Vector Machine (SVM) is a generalized linear binary classifier. SVM aims to find a hyperplane that separates all samples of different types by supervised learning. SVM has a special function that it can map the feature data to the high-dimensional feature space through a kernel function and then find the dividing hyperplane, thereby realizing linear or nonlinear classification. The kernel functions usually used are Linear kernel, Polynomial kernel and Gaussian kernel (equation 2,3,4). SVM has many advantages. The complexity of SVM algorithm is related to the number of support vectors. SVM is not sensitive to the dimension of the feature space, so it is also very effective in high-dimensional space. Besides, SVM is not sensitive to abnormal value, and has good generalization ability. On the other hand, SVM have a long training time and is more sensitive to the selection of kernel functions.

$$\text{Linear Kernel } K(x, z) = x^T z = x \cdot z = <x, z> \tag{2}$$

$$\text{Polynomial kernel } K(x, z) = (ax^T z + c)^d = (ax \cdot z + b)^d \tag{3}$$

$$\text{Gaussian kernel } K(x, z) = e^{-\frac{||x-y||^2}{2\sigma^2}} \tag{4}$$

Multilayer Perceptron (MLP) is a typical forward propagation artificial neural network (ANN) model. MLP consists of an input layer, one or more hidden layers and an output layer, and the layers are fully connected. Except for the input node, all other nodes are neurons with nonlinear activation functions. MLP takes supervised learning by back-propagating the output error. MLP performs well on nonlinear data and is also suitable for multi-classification problems. However, the number of layers of the model and the number of hidden neuron nodes are uncertain. In addition, MLP is slow to train and prone to overfitting, and the interpretability of the model is not strong.

$$y = \sigma(f(x)) = \sigma(w^T x) = \frac{1}{1+e^{-w^T x}} \tag{5}$$

Logistic regression model is a linear classification model that frequently applied to binary classification problems. The model is relatively simple to implement, requires little computation, and has fast training speed. However, logistic regression model has obvious limitations, such as the performance for linear inseparable problems may be very poor, the model is prone to underfitting, and the classification accuracy is not high. Logistic regression model can be described as equation (5).

The K-Nearest Neighbors (KNN) algorithm is a simple ML classification model. For an unknown sample (i.e. waiting to be classified) in a specific space, if most of the K-nearest samples (neighbors) near the unknown sample belong to a certain category, the unknown sample will also be judged to belong to this category. As equation (6) shows. The KNN model makes no assumptions about the data, does not require parameter estimation, and can also be applied to multi-classification problems. However, the KNN model is highly dependent on training data, and when the amount of data or feature space is particularly large, this classification algorithm requires a lot of computation. When the samples are unbalanced, the prediction accuracy of the categories with a small number of samples will be low. Similar to the KNN model, the Nearest Centroid (NC) classification model uses the centroids of different category's members to represent this category respectively, and the sample waiting to be classified will take the category whose centroid is closest. As equation (7) shows.

$$y = \underset{c_j}{\mathrm{argmax}} \sum_{x_i \in N_{k(x)}} I(y_i, c_j), \quad i = 1,2,\ldots,N; j = 1,2,\ldots,K \qquad (6)$$

$$y = \underset{c_j}{\mathrm{argmin}} \{ D(y, c_i) \}, \quad i = 1,2,\ldots,K \qquad (7)$$

Naive Bayes (NB) classifiers are a series of simple probabilistic classifiers based on Bayes theorem, for which a premise is that all features must be independent of each other. Gaussian, Multinomial and Bernoulli are three frequently used NB model. For the sample to be classified, the category with the highest probability is selected to predict its category (classification result), by calculating the probability of occurrence of all different categories under the condition that this sample occurs. As equation (8) shows. NB classifiers are less sensitive to data items missing and can be applied to multi-classification problems. However, NB classifiers need to calculate the prior probability, and are limited by the premise that feature independence, which may affect the accuracy of the classifiers.

$$y = f(x) = \underset{c_k}{\mathrm{argmax}} P(Y = c_k \mid X = x) = \underset{c_k}{\mathrm{argmax}} \frac{P(Y=c_k) \prod_j P(X_j = x_j | Y = c_k)}{\sum_k P(Y=c_k) \prod_j P(X_j = x_j | Y = c_k)} \qquad (8)$$

The Adaptive Boosting (AdaBoost) [24] algorithm improves the classification performance by increasing attention to error-prone samples. AdaBoost first trains a weak learner (decision tree adopted in this paper) from the data; then, iteratively, increases the weight of error-prone samples while decreasing the weight of correctly predicted samples and trains further. In each iteration, the next weak learner is trained on the adjusted weight parameters, and finally a strong classifier is integrated by these weak learners and their weight parameters. This ML model takes advantage of the cascade of weak learners and considers more detailed characteristics of the data, so it has a higher accuracy in theory. On the other hand, the model is relatively difficult to train, and the data's imbalance may lead to a drop in classification accuracy.

The Voting classifier is a typical ensemble classifier. It first obtains multiple ML classifiers or regression models on the same training dataset, and then synthesizes the results of these classifiers to judge the final classification results by majority voting (hard voting) or average prediction probability (soft voting). The Voting model can improve the generalization ability and increase the stability of the classifier. In this paper, five sub-classifiers, i.e. Decision Tree, the Nearest Centroid classifier, Multilayer Perceptron, Support Vector Machine and Naive Bayes classifier are used in the Voting classifier, and the final result is predicted by hard voting.

## 2.4. Threat Detection

After 9 types of ML classifiers described in section 2.3 are trained, they are used to detect general social engineering attack threat, respectively. Since each kind of classifier is trained on 3 different feature combination datasets (as presented in section 2.2), there are finally 27 ML classifiers in total to detect SE attack threats.

## 3. Experiment Result and Analysis

### 3.1. Experiments Method and Evaluation Indicators

For every time of threat detection, the five-fold cross-validation was used to measure the average performance of the ML classifier. In order to further reflect the average performance of these 27 classifiers, for each classifier, this paper conducts 10 times of training and cross-validation, and respectively calculates the average value of the precision, recall and F1-score.

$$\text{precision} = \frac{\text{true positive}}{\text{true positive} + \text{false positive}} = \frac{\text{true positive}}{\text{No. of predicted positive}} \tag{9}$$

$$\text{recall} = \frac{\text{true positive}}{\text{true positive} + \text{false negative}} = \frac{\text{true positive}}{\text{No. of actual positive}} \tag{10}$$

$$\text{F1-score} = 2 \times \frac{\text{precision} \times \text{recall}}{\text{precision} + \text{recall}} \tag{11}$$

In this paper, the inherent precision, recall and F1-score of the positive samples in the dataset, namely 0.613, are used as the comparison baseline for ML threat detectors (classifiers).

### 3.2. Experimental Results

Table 3 summarizes these experimental results. Figure 5, Figure 6 and Figure 7 are the boxplots of the precision, recall and F1-scores of the 9 types of classifiers for 10 times of experiments with 3 different feature combinations.

Table 3 The average performance of SE threat detection classifiers using different kinds of ML models

| FC No. / Criteria / ML models | Feature Combination 1 | | | Feature Combination 2 | | | Feature Combination 3 | | |
|---|---|---|---|---|---|---|---|---|---|
| | Precision | Recall | F1 | Precision | Recall | F1 | Precision | Recall | F1 |
| Decision tree | 0.742 | 0.761 | 0.749 | 0.869 | 0.848 | 0.857 | 0.896 | 0.855 | 0.876 |
| Random forest | 0.714 | 0.808 | 0.758 | 0.746 | 0.816 | 0.778 | 0.78 | 0.825 | 0.802 |
| SVM | 0.708 | 0.789 | 0.746 | 0.695 | 0.738 | 0.716 | 0.78 | 0.825 | 0.802 |
| MLP | 0.734 | 0.604 | 0.659 | 0.73 | 0.778 | 0.752 | 0.786 | 0.726 | 0.753 |
| Logistic regression | 0.639 | 0.629 | 0.633 | 0.657 | 0.691 | 0.672 | 0.692 | 0.717 | 0.704 |
| Nearest Centroid | 0.766 | 0.777 | 0.77 | 0.715 | 0.759 | 0.736 | 0.747 | 0.771 | 0.759 |
| Naive Bayes | 0.715 | 0.709 | 0.711 | 0.698 | 0.629 | 0.663 | 0.741 | 0.701 | 0.72 |
| Adaptive Boosting | 0.719 | 0.785 | 0.75 | 0.723 | 0.799 | 0.757 | 0.715 | 0.813 | 0.761 |
| Voting | 0.742 | 0.787 | 0.763 | 0.771 | 0.773 | 0.77 | 0.793 | 0.791 | 0.791 |

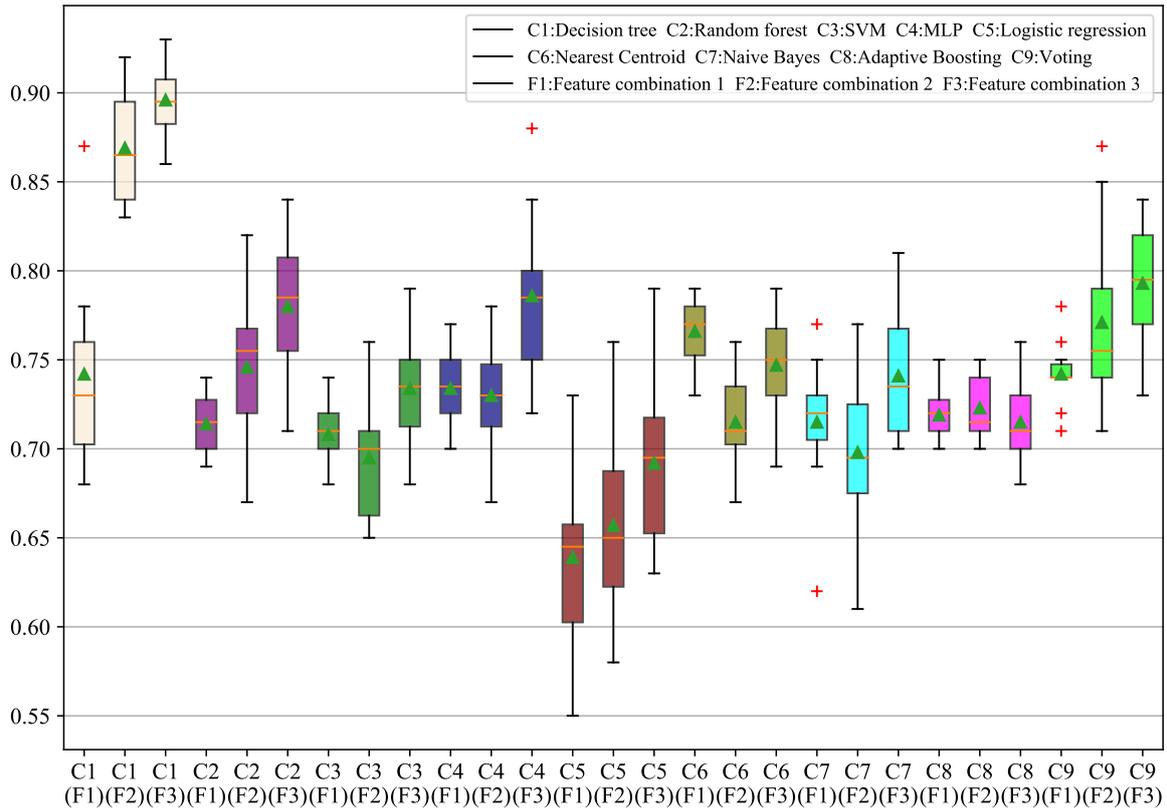

Figure 5: The precision of 9 classifiers to detect SE threat on different feature combinations

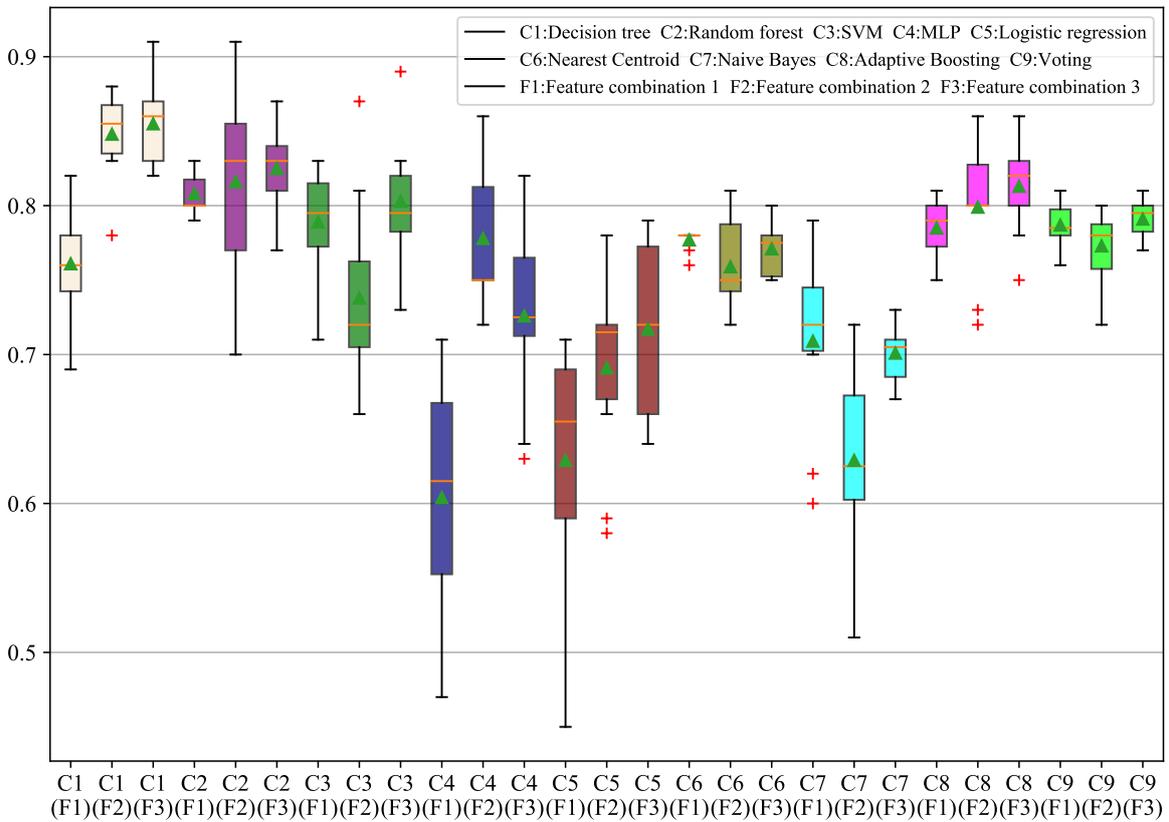

Figure 6: The recall of 9 classifiers to detect SE threat on different feature combinations

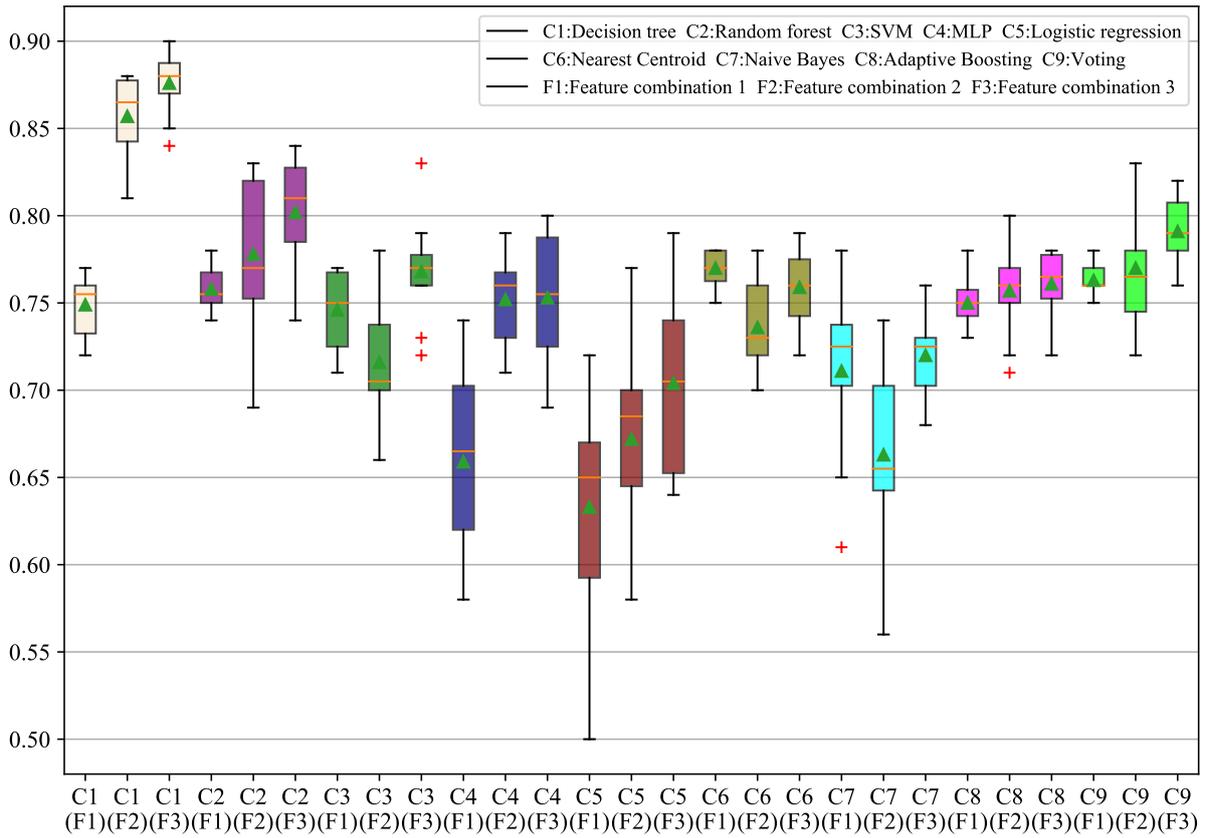

Figure 7: The F1 score of 9 classifiers to detect SE threat on different feature combinations

### 3.3. Analyses and Discussions

Comparing the work [22] (knowledge graph-based SE threat analysis and detection) and SE threat detection using ML in this paper, as well as the experimental results among different SE threat detection classifiers using ML, the following conclusions and analyses can be drawn:

(1) As the number of features increases, the comprehensive performance (F1-score) of the 9 classifiers improves on the whole (Figure 7). For feature combination 3, the average F1-score over ten times of experiments for all 9 classifiers exceed 70%. Seven of these classifiers had F1-scores over 75%, three over 80%, and one over 85% (Table 3). These experimental results demonstrate the feasibility for threat detection for general SE attacks using ML techniques , as well as the usability of the ML dataset generated in this paper.

(2) *1)* When the feature number is small (feature combination 1), the SE threat detection classifiers (detector) that use Decision Tree, Random Forests, the Nearest Centroid and AdaBoost perform better, all of which have F1-score above 75%. *2)* As the feature number increases, threat detector using Decision Tree, Random Forests, SVM, MLP and AdaBoost perform better, with higher precision and F1-score (Table 3). And the performance of the detectors using Decision Tree, Random Forests, and AdaBoost improved steadily and significantly (Table 3, Figure 7). *3)* Regardless of the feature number, the performance of threat detectors using Logistic Regression and Naive Bayes is comparatively poor. It can be inferred that the SE dataset in this paper is likely to conform to certain rules, which is suitable for tree types, rules-learned and/or some sophisticated ML model; simple linear classifiers and Naive Bayesian classifiers are limited in the practice.

(3) Compared to other threat detectors, the overall performance of Decision Tree is the best. When only two features are considered (feature combination 1), the average precision, recall, and F1-score of the decision tree are 74.2%, 76.1%, and 74.9%, respectively. As the number of features increases, the performance of the decision tree improves steadily. When considering four features (feature combination 3), the average precision, recall and F1-score of the Decision Tree respectively reach 89.6%, 85.5% and 87.6%, which shows the effectiveness of feature extraction and fusion (section 2.2) and the trained Decision Tree SE threat detector.

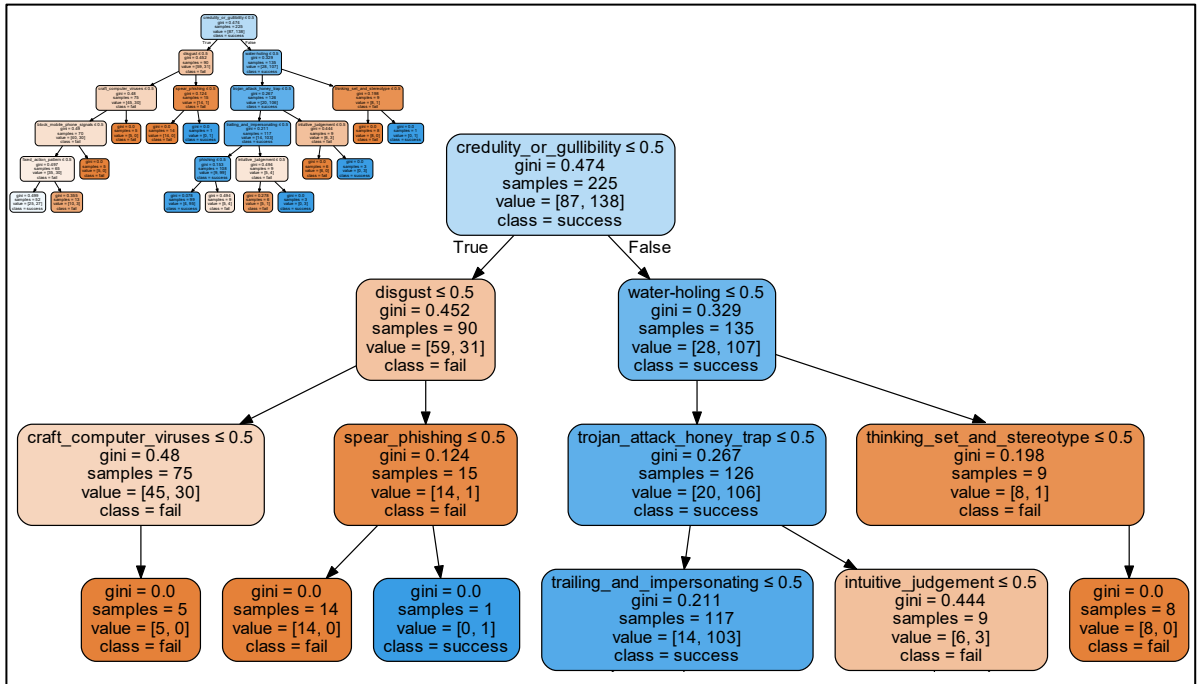

Figure 8: The decision tree to detect social engineering threat correspond with feature combination 1

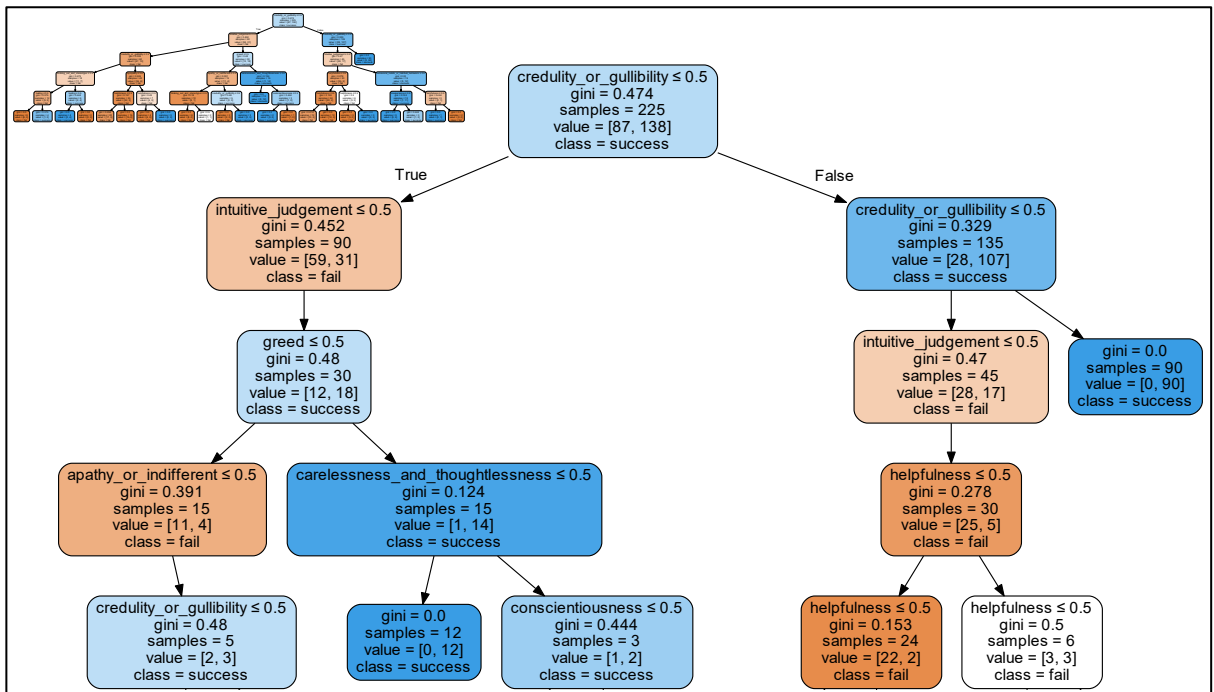

Figure 9: The decision tree to detect social engineering threat correspond with feature combination 2

(4) After adding the vulnerability feature of SE attack method (feature 3), the average precision, recall and F1-score of the decision tree were significantly improved, reaching 86.9%, 84.8%, and 85.7%, respectively. By comparing the contents of the Decision Tree 1 and 2 respectively trained by the dataset 1 and 2 (correspond with feature combination 1 and 2) in the ten experiments, it is found that, the number of attack method nodes and human vulnerability nodes in decision tree 1 is basically the same, as shown in Figure 8; the human vulnerability nodes in decision tree 2 accounts for the vast majority, and only a small number of attack method nodes appear at the bottom of the tree, as shown in Figure 9. This reflects the significant impact of feature 3 on the performance of Decision Tree SE threat detector. Besides, in the ten experiments for the two feature combination datasets, the root nodes of all the Decision Trees are "credulity_or_gullibility" (similar to Figure 8 and Figure 9), which is exactly the most exploited human vulnerability detected by the knowledge graph-based approach in [22]. These results suggest that human vulnerability is critical to SE threat prediction.

(5) From the structure and content of the Decision Trees in Figure 8 and Figure 9, it can be seen that the Decision Trees did not or cannot learn the abstract logical analysis patterns or rules like work [22]. Instead, Decision Trees acquired the statistical knowledge (and built a classification model) from the feature dataset to judge whether an attacker can succeed in posing a SE threat to a victim. This reflects the essential difference between SE threat detection techniques using machine learning and approaches based on knowledge graph presented in work [22], and enriches the threat detection technologies for SE attacks.

(6) The performance of the Nearest Centroid classifier is relatively stable, and the performance drops slightly after adding feature 3. This may imply that the feature 3 reduces the distance between the centroids of different categories of samples in the original feature space and increases the misclassification probability for test samples.

(7) The precision of the Logistic Regression classifier that correspond with feature combination 1 is only 63.9%, which is close to the baseline of positive samples (61.3%). Although the three performance indicators of the Logistic Regression classifier improved with the increase of features, the overall performance is still relatively poor. Two reasons may account for the under-fitting of the model: 1) the number of threat features considered is not enough; 2) it is a linear inseparable problem in these datasets to determine whether SE threat exists in a pair of <attacker, target>.

(8) After adding feature 3 (feature combination 2), the performance of the Naive Bayes classifier dropped significantly. This is because a linear correlation exists between feature vector 3 and feature vector 2, which violates the independence premise that the Naive Bayes classifier relies on, thus affecting the classification.

(9) After adding feature 3 (feature combination 2), the performance of SVM decreases instead. This is because the SVM model is very sensitive to the choice of kernel function. The SVM created in the experiments uses a Gaussian kernel (radial basis) function, which is mainly used to deal with the linearly inseparable situation. However, feature vector 3 has a linear correlation with feature vector 2, which negatively affects the model training. In other words, SVM maps the originally separable data to a high-dimensional space and then attempts to find a dividing hyperplane, which increases the possibility of misclassification and leads to performance degradation. With the addition of non-linear feature 4, the model learned new statistical features, and the classification performance improved.

(10) Despite being negatively affected or undulated by the Naive Bayes, SVM and the Nearest Centroid classifiers, the Voting classifier reduces the prediction error and steadily improves the detection performance by the advantages of the voting mechanism and the performance improvement of Decision Tree and MLP. This shows that the voting model has a better the stability and generalization ability.

(11) With the increase of the feature number, the overall performance of Random Forests, MLP neural networks and AdaBoost classifiers steadily improved (Figure 6, Figure 7), indicating that increasing the number of SE threat features is conducive to the training of complex machine learning models. However, after vectorizing the effect mechanism feature of human vulnerability (feature vector 5) and adding it to feature combination 3 (i.e. forming a 164-dimensional feature vector, called feature combination 4), it was found that the performance of the new trained 9 types of classifiers remained to the current level. This shows that the effective feature gain introduced by the new feature vector is basically equal to the noise cost introduced at the same time.

(12) Combining conclusion (1) - (11), especially (1)(2)(3)(4)(5)(11), considering the original dataset (section 2.1, work [22]) and feature extraction and fusion (section 2.2), it can be found that the SE domain ontology [22] can effectively dissect SE attacks and model the inherent SE threat features, and it therefore can be used as a data model to build larger SE datasets used for more security studies on social engineering attacks.

Besides, although ML-based SE threat detection techniques are not comparable to the graph-based approaches in the visualization and flexibility, the ML classifiers automatically establish the classification models by fitting the statistical characteristics of the training dataset, don't limit to learn the logic patterns presented in work [22] and don't rely on the path accessibility or sub-graph retrieval in the knowledge graph. Thus, SE threat detection using ML techniques is complementary with graph-based approaches and has a good application significance.

(13) About the stability of the classifiers' performance. 1) In the three boxplots, there are a few classifiers with stable performance. For some classifiers, the distance between the maximum and minimum of their performance indicators is large, the box body (Q3-Q1) is long, and many outliers appear in the boxplot of recall performance. These results indicate that the performance of certain classifiers is not stable enough. 2) Random Forest, AdaBoost and MLP are also three classifiers whose performances are relatively stable and growing, although their performances are not as good as Decision Tree. This paper tried to adjust the parameters of these three classifiers, but the overall performance did not improve significantly. One reason may be that complex models are difficult to train, and another reason may be the datasets used in this paper are relatively small.

Nonetheless, the experimental results (Table 3, Figure 5, Figure 6, Figure 7) can be obtained on the dataset with only 225 labels, indicating that the stability of the classifiers is acceptable. The difficulty to get datasets for general SE attacks has been discussed in work [22], and we will research on SE threat detection on real network and big data environment in the future.

## 4. Conclusion

This paper focused on the threat detection for general social engineering attack using machine learning techniques. We got more threat data from the knowledge graph presented in [22] at first, and then generated new training & testing datasets correspond with 3 different feature combinations after feature extraction and fusion. Finally, 9 types of ML models were created, and 27 threat detectors/classifiers were trained and tested ten times, respectively.

The experimental results and analysis suggest that, the threat detection using ML techniques for general SE attack are feasible and effective, and it is complementary with graph-based approaches. For the datasets in this paper, the best ML detection model is Decision Tree; its average precision, recall and F1-score of reach 89.6%, 85.5% and 87.6% respectively when considering four threat features; Random Forest, SVM, MLP, the Nearest Centroid, AdaBoost and integrated Voting models also perform well. The experiments also validate the usability of these datasets, demonstrating that the SE domain ontology [22] can dissect SE attacks and deliver the SE threat features, and it therefore can be used as a data model for future research. Other discussion about the ML models, datasets, detection performance, etc. are also presented.

## Data Availability

The data and materials of this study are available from the corresponding author upon reasonable request.